\documentclass[11pt]{article}
\usepackage{graphicx} 
\usepackage[utf8]{inputenc}
\usepackage[backend=biber,style=apa]{biblatex} 
\usepackage[margin=1in]{geometry}
\usepackage[T1]{fontenc}
\usepackage{caption}
\DeclareCaptionFormat{apatable}{%
  \textbf{#1#2}\par
  \vspace{\baselineskip}
  \textit{#3}\par
  \vspace{\baselineskip}
}

\usepackage{siunitx}
\usepackage{threeparttable}
\usepackage{lmodern}
\usepackage{microtype}
\usepackage{float}
\usepackage{graphicx}
\usepackage{makecell}
\usepackage{booktabs}
\usepackage{longtable}
\usepackage{enumitem}
\usepackage{amsmath, amssymb}
\usepackage[hidelinks]{hyperref}
\usepackage{xcolor}

\begin{document}
\renewcommand{\thefootnote}{\fnsymbol{footnote}}
\begin{center}

\rule{\linewidth}{0.8pt}\\[0.8em]
{\large\scshape Estimating Item Difficulty with Large Language Models as Experts}\\[0.8em]
\rule{\linewidth}{0.8pt}

\bigskip\bigskip

\textbf{Diana Kolesnikova}$^{1}$\footnotemark \quad
\textbf{Kirill Fedyanin}$^{2}$ \quad
\textbf{Abe D. Hofman}$^{3,4}$\\[0.5em]
\textbf{Matthieu J. S. Brinkhuis}$^{5}$ \quad
\textbf{Maria Bolsinova}$^{1}$

\bigskip

{\small
$^{1}$Department of Methodology and Statistics, Tilburg University, Tilburg, Netherlands\\[0.25em]
$^{2}$Smart Business Technologies, Belgrade, Serbia\\[0.25em]
$^{3}$Department of Psychological Methods, University of Amsterdam, Amsterdam, Netherlands\\[0.25em]
$^{4}$Prowise Learn, Amsterdam, Netherlands\\[0.25em]
$^{5}$Department of Information and Computing Sciences, Utrecht University, Utrecht, Netherlands
}

\bigskip

{April 17, 2026}

\end{center}

\footnotetext{Corresponding author: Diana Kolesnikova,
\href{mailto:d.kolesnikova@tilburguniversity.edu}{\texttt{d.kolesnikova@tilburguniversity.edu}}}

\section*{Abstract}
Accurate estimates of item difficulty are essential for valid assessment and effective adaptive learning. However, for newly created tasks, response data are typically unavailable. Pretesting and expert judgement can be costly and slow, while machine learning methods often require large labelled training datasets. Recent work suggests that large language models (LLMs) may help. However, there is limited evidence on the elicitation procedures and prompt configurations used to emulate experts for difficulty estimation. This study addresses this gap by evaluating three off-the-shelf LLMs as difficulty raters for newly created items without access to response data.

Using an item bank from an online learning system, the study examined 6 domains of primary-school mathematics, with empirical difficulty estimates treated as empirical reference. The study used a full factorial design crossing three factors: judgement format (absolute vs pairwise), decision type (hard decisions vs token-probability-based estimates), and prompting strategy (zero-shot vs few-shot). LLM-derived difficulty estimates were compared with empirical difficulties using Spearman rank correlations. 

Across domains, LLM-based estimates exhibited moderate to strong positive correlations with empirical item difficulties. For simpler arithmetic tasks, some configurations approached the upper end of the accuracy range reported for human experts in previous research. Pairwise comparison consistently outperformed absolute judgement in the absence of additional refinements. However, when token-level probabilities were incorporated and examples of items with known empirical difficulty were provided, the absolute judgement configuration likewise demonstrated moderate-to-high alignment. 

By systematically comparing key design choices, the study positions LLMs as a promising tool for initial item calibration and offers insights into effective workflow configuration in settings where many new items are generated and response data are unavailable.

\section*{Keywords} 
Educational Assessment, Task Difficulty Estimation, Automated Item Calibration, Simulating Expert Behavior, Standard Setting, Pairwise Comparison, LLM-as-a-judge

\section*{Practitioner notes}

\textbf{What is already known about this topic}
\begin{itemize}
    \item In both assessment and learning contexts, newly developed items must be calibrated, typically through pretesting or expert judgement.
    \item The quality of expert judgement varies considerably and evidence comparing specific procedures remains limited. In particular, pairwise judgement is often claimed to outperform absolute judgement, but the evidence is still inconclusive.
    \item LLMs are increasingly used to generate synthetic data, for example, simulated student responses, yet only a few studies have examined their ability to emulate expert judgement for item calibration.
\end{itemize}

\textbf{What this paper adds}
\begin{itemize}
    \item This study investigates whether directly prompted LLMs can emulate expert judgement for item difficulty estimation.
    \item A systematic comparison of key design choices is conducted: (i) pairwise versus absolute procedures; (ii) the use of token probabilities versus hard-decision outputs; (iii) the use of examples in prompts (few-shot) versus a zero-shot strategy.
    \item The comparison is based on six domains of primary-school mathematics, enabling the evaluation of robustness across task domains.
     \item Configurations that improve the quality of the estimates are described.
\end{itemize}

\textbf{Implications for practice and/or policy}
\begin{itemize}
    \item LLMs have the potential to serve as a cheaper and faster tool for item difficulty estimation, especially when large-scale pretesting is infeasible, including in automated item generation workflows where many new items must be evaluated before response data are available.
    \item Pairwise comparison appears to be the most robust LLM-based strategy in terms of alignment with empirical estimates. However, because pairwise procedures are generally more time- and cost-intensive, absolute judgement may offer a more scalable alternative when combined with token probabilities and few-shot prompting.
\end{itemize}

\section{Introduction}

Reliable estimates of item difficulty are essential for valid assessments \parencite{WainerMislevy2000} and efficient learning systems \parencite{Wauters2012}. Knowledge of task difficulty supports score comparability in high-stakes examinations, enables the construction of parallel test forms, and targeted item revision  \parencite{Benton2020, VanderLindenAdema1998}. In adaptive learning systems, it helps educators mitigate the cold-start problem \parencite{VeldeEtAl2021, Klinkenberg2011}, and design efficient learning trajectories \parencite{KulikFletcher2016, VanLehn2011}.

Difficulty is typically conceptualised as a latent characteristic of an item that influences the probability of a correct response (e.g., \cite{HambletonSwaminathanRogers1991}). In practice, difficulty is commonly quantified through proportions correct in Classical Test Theory (CTT) methods (\cite{CrockerAlgina1986}) or item parameters in Item Response Theory (IRT) models \parencite{LordNovick1968, Rasch1960}, however, often data on actual student responses are unavailable.

In these cases, difficulty is estimated through pretesting or expert judgement. Pretesting provides direct evidence, but can be costly, slow, and increase the risks of items' exposure \parencite{JaliliHejriNorcini2011,JaliliHejriNorcini2011,OzakiToyoda2006}. Expert judgement requires extensive training \parencite{LivingstonZieky1982, HambletonEtAl2003} and often exhibits substantial variability in accuracy (correlation between expert estimates and empirically obtained difficulties $r \approx  [.0, .9]$; \cite{Attali2014}).

Recent years have seen growing interest in automated methods for difficulty estimation, particularly machine-learning (ML) approaches \parencite{AlKhuzaey2024}. Supervised ML models can be highly accurate but require labelled datasets \parencite{Benedetto2020March}. Methods based on large language models (LLMs) have been explored to simulate student responses; however, a few published works yield unrealistic ability distributions \parencite{LiuBhandariPardos2025}, rely on large model ensembles (e.g., 65 LLMs in \cite{ParkEtAl2024}), and most approaches still depend on item response records to fine-tune LLMs, which limits their usefulness in cold-start settings.

To address the limitations mentioned above, the present study evaluates the capacity of three off-the-shelf LLMs to simulate expert difficulty judgements without relying on extensive training data. The study addresses five research questions:
\begin{itemize}

\item RQ1 Can off-the-shelf LLMs approach the upper bound of difficulty-estimation quality reported in the literature for human raters?
\end{itemize}

\begin{itemize} 
\item RQ2 Does the pairwise comparison procedure, as opposed to the absolute judgement, improve the alignment between LLM-derived and empirical item difficulty estimates?
\end{itemize}

\begin{itemize}
\item RQ3 Does token-level probability information ('soft' outputs) improve the quality of LLM difficulty estimates compared to 'hard', single-decision, outputs?
\end{itemize}

\begin{itemize}
\item RQ4 How does the few-shot prompting affect the quality of LLM difficulty estimates compared to a zero-shot strategy?
\end{itemize} 

\begin{itemize}
\item RQ5 How much does the quality of the LLM difficulty estimates vary across different domains?
\end{itemize}

The evaluation used an item bank from an online learning system MathGarden\footnote{Part of Prowise Learn; see \url{https://www.prowise.com/en/product/prowise-learn}} \parencite{Klinkenberg2011, Straatemeier2014, Brinkhuis2018},
 covering six primary-school mathematics domains. Within each domain, 60 items were sampled, and their LLM-derived difficulties were compared to the empirical difficulty parameters using bootstrapped intervals for Spearman rank correlation.

This study makes two contributions. (i) First, it provides a systematic comparison of LLM-based difficulty estimation across judgement format, prompting strategy, and decision type. (ii) Second, it offers practical guidance, including R scripts and prompts \footnote{\url{https://osf.io/fpa5v/overview?view_only=344f334178c74b36883f8e59ed5fe42a}}, for configuring LLM-as-a-judge workflows for preliminary item calibration, which could be valuable for future research and applied use.

The remainder of this paper is structured as follows. Section 2 reviews the related literature on difficulty estimation methods and the LLM-as-a-judge research. Section 3 outlines the methodology of the study. Section 4 reports the results. Section 5 discusses limitations and possible ideas for future research.

\section{Related work}

 This section reviews two groups of studies devoted to difficulty estimation. First, it summarises procedures based on human expert judgement, and second, it reviews automated approaches, from early ML methods to recent uses of LLMs for simulating students and experts. The section concludes by identifying the gaps that motivate the comparisons conducted in the present study.

\subsection{Human expert procedures}

Research on calibrating newly developed items using human experts typically follows two related strands. One line of work asks experts to estimate item difficulty for the overall target population (expected proportion correct; \cite{Gulliksen1950, Thorndike1982, ImparaPlake1998}). The second line of work emerges from standard-setting research, where experts judge the item performance for a minimally competent (borderline) candidate as part of the establishment of cut scores \parencite{Angoff1971, Attali2014, Nedelsky1954}. Although the target differs, both branches address the same practical problem: generating plausible item-difficulty information before large-scale administration.

Across studies, the accuracy of expert difficulty judgements is mixed, ranging from strong alignment with empirical difficulty estimates to weak or even zero relationship. For total-group estimation, reported correlations between expert judgements and empirical difficulty can be high when aggregated across judges (e.g., Pearson $r \approx .72$--$.83$ in \cite{Thorndike1982}; $r \approx .78$ in \cite{ImparaPlake1998}), yet individual-level performance varies substantially (e.g., Pearson $r \approx .40$--$.85$ across raters in \cite{LorgeKruglov1953}). Other investigations report considerably lower alignment even with extensive training; for example, \textcite{Bejar1983} found correlations below .50 between group estimates of a delta difficulty index and empirical estimates for test items. In standard-setting contexts, similar variability is observed: correlations between judged and empirical difficulties have been reported from close to zero values up to moderately high values (e.g., roughly $r \approx .04$ - .54 in \cite{MelicanMillsPlake1989}; $r \approx .03$ - $.80$ for individual judges in \cite{Cross1984}). These findings suggest that experts can provide informative signalsb but the strength of that signals depends heavily on study design and the aggregation procedures.

Expert judgement is not a single procedure, but a family of methods that differ along multiple design parameters, including the amount and structure of training (\cite{ImparaPlake1998}; \cite{LivingstonZieky1982}), the number of judges and the aggregation strategy \parencite{Thorndike1982}, whether judges are required to communicate and reconcile discrepancies \parencite{Bejar1983, Clauser2008}, and what response format is required. The current study does not implement any training or communication between LLM-judges, only two design parameters are varied: judgement format and auxiliary information. These parameters are reviewed below.

Common judgement formats include (i) absolute judgements \parencite{ImparaPlake1998}; (ii) ranking items \parencite{LorgeKruglov1952,LorgeKruglov1953,LorgeDiamond1954, Attali2014}, (iii) classifying items into different categories or evaluating them using Likert scale \parencite{Metin2017, Ryan1968, WautersDesmetVandenNoortgate2011EDM, Thorndike1982} or (iv) some combination of the above methods \parencite{Tinkelman1947}.

Comparative judgement approaches, including ranking, have often been reported to outperform absolute judgement procedures \parencite{Attali2014, Benton2020}. However, there is no direct comparison of the two procedures implemented independently using the same quality metric. For example, \textcite{Tinkelman1947} asked experts to estimate the absolute item difficulty but then suggested putting them on one scale and resolving any discrepancies that appear. \textcite{LorgeKruglov1953} suggested the same actions in the opposite order: to rank items and then to estimate their absolute difficulty. Afterwards, they evaluate the quality of ranking using Spearman correlation (pooled judgements and empirical rank varies from .780 to .839; for individual raters the range is .515-.838) and for absolute difficulty (“percent passing”) they obtained Pearson correlation (for mean estimate and empirical value the range is .797-.874; for individual raters .397-.847). In their article of 1952, they claim that relative difficulties are valid estimates based on high Spearman correlation (.83-.84) but absolute estimates differ significantly from true ones based on t-test. Taken together, these studies suggest that the perceived superiority of relative over absolute judgement may rest on the more consistent correlations reported in comparative judgement studies, whereas findings from absolute judgement studies appear more variable and sometimes show significant discrepancies in absolute values.

Within comparative judgement designs, the number of items presented for comparison can vary. It could be the total number of items in the exam, e.g., 45 items \parencite{LorgeDiamond1954, LorgeKruglov1953} or it could be a smaller subset or numerous subsets of fewer items, e.g. 7 items \parencite{Attali2014} or 4 \parencite{CurcinBlackBramley2009}. The simplest case considered in the present study is the pairwise comparison \parencite{Bramley2010, Choppin1968, Ofqual2015}, the procedure rooted in the work of \textcite{Thurstone1927} on the comparison of two stimuli.

Another parameter that is varied in this paper is the availability of empirical characteristics for some of the items ('anchor items', \cite{Thorndike1982}). The evidence on the effect of adding examples or providing additional information is mixed. \textcite{LorgeKruglov1952} provided information on 30 items of 150 items to half of their judges (4 doctorate students) and did not find significant differences between two groups of judges. Later, \textcite{LorgeKruglov1953} increased the number of experts and changed the procedure. Pearson correlations for mean expert judgement and empirical percentage passing were .797-.844 for judges without information and slightly higher .853-.874 for those with extra information. On the other hand, \textcite{Clauser2008} reported that, for the subset of items for which the performance information of the examinee was provided, the correlation between the mean Angoff judgements and the empirical conditional proportion correct increased from .34 (post-discussion) to .66 after review of the performance data. Similar improvements were reported by \textcite{CizekFitzgerald1996}.

Previous work indicates that expert judgements can approximate empirical difficulty, but with substantial variability between studies. This variability has been attributed to several factors, including (i) judgement format, (ii) the availability of calibration information, and (iii) the degree of aggregation and procedural support (e.g., training and discussion). In LLM-based expert simulation, the first two factors can be systematically manipulated through prompt design, whereas the third is more difficult to implement. Accordingly, the present study focuses on the judgement format and calibration information, and holds procedural support constant rather than treating it as an experimental factor.

\subsection{Automated judgement}

Before widespread adoption of LLMs, the main alternatives to pretesting and expert judgement were ML approaches that required labelled datasets (\cite{Benedetto2020March}; \cite{Feng2025}). Feature-based methods (e.g., linear regression, random forests, gradient boosting) relied on engineered linguistic descriptors ranging from simple surface features such as word length \parencite{PandarovaEtAl2019} or the number of operators for math problems \parencite{Feng2025} to more elaborate lexical, syntactic, semantic measures, e.g., readability, semantic ambiguity, text cohesion, psycholinguistic properties \parencite{YanevaBaldwinMee2019}. Subsequent work reduced the dependence on manually specified descriptors by adopting higher-dimensional text representations, including Term Frequency-Inverse Document Frequency (TF–IDF) vectors and static word embeddings aggregated at the item level \parencite{Benedetto2020March, Benedetto2020Jun, YanevaBaldwinMee2019}. 

Some of these studies combine ML and psychometric approaches, e.g., BERT-IRT \parencite{YanceyEtAl2024} which achieved associations of .88-.89 for unseen items similar to the training dataset and .76 for new word stems; or the approach of \textcite{Zoucha2025} where they combined the concept of item characteristic curves with neural nets and achieved accuracy of prediction for item difficulty category equal to 79.6\%). However, these representations remain largely context-independent, which motivated the shift to pretrained transformer encoders that map item text to contextual embeddings and can be used either as fixed feature extractors or fine-tuned for difficulty prediction (\cite{Benedetto2021Apr}; \cite{LiEtAl2025}). 

Although evidence suggests that such ML approaches can perform comparably to, or somewhat better than, human experts (RMSE values for continuous difficulty prediction ranging from 0.666 to 0.978; compared to 1.004 for human experts; for a five-category classification task, predictive accuracy from 0.425 to 0.650 for ML methods, with human accuracy at 0.650; \cite{Stepanek2023}), these models typically required domain specific training sets ranging from roughly hundreds to thousands of items (depending on feature dimensionality), which makes these methods applicable only to certain contexts. Increases in model and training scale, the adoption of instruction tuning, and, crucially, the shift from encoder-only architectures to decoder-only generative models have enabled direct interaction via prompting, making contemporary LLMs a distinct tool for item difficulty
prediction.

One way to employ LLMs with direct prompting is to simulate possible student response patterns and then apply IRT or other psychometric models to derive difficulties \parencite{SrivatsaEtAl2025,Marquez2025, Acquaye2026,LiEtAl2025StudentStruggles}. One reported limitation is the
narrow distribution of simulated student abilities \parencite{LiuBhandariPardos2025}. To overcome this limitation, multiple LLMs have been used. For example, \textcite{ParkEtAl2024} employed 65 different LLMs from different generations and combined them into a few groups, so that the answer patterns could resemble the answers of the weaker and stronger students. This approach used student response records, however, they could be replaced by other sources of information about their abilities, e.g., grades or teacher judgements.

An alternative to student-simulation approaches is to elicit expert-like judgements directly from LLMs. In the broader computer science literature, this paradigm is commonly referred to as LLM-as-a-judge, reflecting the use of LLMs as evaluators for complex outputs \parencite{ZhengEtAl2023}.

Judgements can be elicited in either an absolute (pointwise) or comparative (pairwise/ranking) format. Both formats have been used on a range of evaluation tasks (e.g., text summarisation and instruction-following evaluation), with mixed evidence regarding their relative advantages \parencite{LiusieEtAl2024PoE, TripathiEtAl2025}. In the specific context of item calibration, existing studies have predominantly employed absolute judgement formats \parencite{Yoshida2024}. For instance, \textcite{Zotos2025} report a relatively weak alignment between direct LLM prompting and empirical difficulty (Spearman $r \approx .052 -.345$), although these estimates still outperformed university professors for items from a Neural Networks exam and an Advanced Machine Learning course. \textcite{Razavi2025} used direct prompting to obtain difficulty predictions that were subsequently combined using regression-derived weights fitted on a training sample ($r \approx .81 -.83$); they reported a feature-based alternative in which task features were used as inputs to random forest ($r \approx .87$) and gradient boosting models. The only study using pairwise comparison procedures reports low-to-moderate correlations between human raters'estimates and pairwise-derived ratings (Spearman $r \approx .16 -.56$; Pearson $r \approx .24 - .53$; \cite{Ballon2025}).

Few-shot prompting is known to improve LLM performance in general by 5-10\% \parencite{Brown2020}. In item calibration, the evidence for the benefits of few-shot prompting is limited but suggestive \parencite{ParkEtAl2024}. \textcite{Zotos2025} include a labelled example in the prompt but do not report a comparison against a zero-shot condition.

Using soft probabilities (i.e., token-level probabilities) can improve performance and reduce resource requirements. \textcite{RainaLiusieGales2025} reported an improvement in Spearman correlation from .41 to .48 when soft probabilities were used to fine-tune LLM to judge reading comprehension tasks' difficulty. \textcite{Zotos2025} similarly showed that incorporating the probability of the first token generated as a feature in supervised
learning improves performance relative to direct prompting and simple text features.

Overall, automated difficulty estimation has progressed from feature-engineered supervised ML to transformer-based representations and, more recently, to prompt-based LLM methods. However, pre-LLM approaches and many LLM student-simulation pipelines remain constrained by the availability of labelled datasets, which limits their usefulness in practice. LLM-as-a-judge approaches have so far shown moderate performance, leaving substantial scope to explore how judgement format (including comparative procedures), few-shot prompting, and the use of soft probabilities can improve difficulty estimation, which is the focus of the current study.

\section{Methodology}

This section outlines the research pipeline and explains the key methodological choices of the study. It begins with a description of the dataset used, followed by a motivation of the selected LLMs. Next, LLM-based procedures for item evaluation are introduced, along with practical challenges encountered during their implementation. The section concludes with strategies for raw data processing and computing outcome measures.

\subsection{Data pre-processing and empirical difficulty estimates}

The dataset was obtained from Prowise Learn and covers the period from 1 September 2017 to 1 July 2018. It contains 8,597 tasks in total, within 6 games in elementary school mathematics. These games are here referred to as domains. They include tasks on addition, subtraction, multiplication, division, calculation order, and text problems. Examples of the tasks are provided in Appendix A. 

To ensure that the empirical difficulties were estimated with enough precision, within each domain only items with the number of responses larger than 300 were retained and then only users who solved more than 5 tasks and fewer than 5,000 tasks were filtered. To maintain independence between domains, only data from one distinct grade level were retained for each domain. Table~\ref{tab:dataset} summarises the resulting dataset by domain prior to difficulty estimation. All pre-processing and analysis was performed in R \parencite{coreteam}.

\begin{table}[ht]
\centering
\begin{threeparttable}
\caption{\textit{Dataset Summary by Domain Prior to Difficulty Estimation}}
\label{tab:dataset}
\begin{tabular}{lrrrr}
\toprule
Domain & Grade & Unique Students & Unique Items & Average Responses per Item \\
\midrule
Addition          & 3 & 23,241 & 882  & 3,537 \\
Subtraction       & 4 & 27,235 & 1,014 & 2,320 \\
Multiplication    & 6 & 23,300 & 729  & 2,379 \\
Division          & 7 & 20,304 & 962  & 1,733 \\
Calculation order & 8 & 12,885 & 720  & 904 \\
Text problems     & 5 & 23,327 & 690  & 4,069 \\
\bottomrule
\end{tabular}

\begin{tablenotes}[flushleft]
\footnotesize
\item \textit{Note}. Grade 3 means that students are 6-7 years old. 
\end{tablenotes}
\end{threeparttable}
\end{table}

Empirical item difficulties that were treated as a criterion in this paper (similar to \cite{AlKhuzaey2024, Benedetto2020March}) were estimated by fitting a Rasch model (\textcite{Rasch1960}) separately within each content domain and then transformed into expected proportions correct. Proportion correct dignifies the proportion of student population who is expected to solve the task correctly. A detailed description of the difficulty-estimation procedure is provided in Appendix B. 

On average, tasks in this dataset exhibited a relatively restricted range of empirical difficulty, with the majority falling at the easier end of the scale. To ensure representation across the difficulty distribution, items for the study were sampled with the stratified procedure. After estimating difficulty, the
item pool was divided into four strata with 25th, 50th and 75th percentiles for item difficulties as split borders. Within each strata, 15 items were sampled without replacement, producing 60 items per domain. This sample size reflects a practical trade-off between achieving acceptable precision for correlation-based analyses and maintaining the feasibility of the pairwise comparison procedure; with 60 items, the design entails 1,770 direct pairwise comparisons. Only open-ended tasks were sampled, as this allowed one to focus on problem wording and not on the format of answer options. For the few-shot strategy, two items were additionally sampled for each domain: one from the easiest and one from the most difficult stratum.

\subsection{Large language models Selection}

Three LLMs were selected for this study. The selection criteria were: (a) strong performance in the Chatbot Arena in the beginning of 2025 \parencite{ZhengEtAl2023}; (b) inclusion of open-source and proprietary models; (c) representation of different model families; (d) availability via an API compatible with the OpenAI protocol; (e) support for retrieving token-level log-probability distributions; and (f) cost considerations. Based on these criteria, GPT-4o (gpt-4o-2024-08-06), DeepSeek-Chat (DeepSeek-V3.2 Release 2026/01/12) and Qwen (qwen3-235b-a22b-instruct-2507) were chosen. Although Qwen is open source, the Alibaba Cloud–hosted deployment is used to simplify implementation. All LLMs were accessed through R script developed by the authors using the package httr \parencite{wickham2023}.
Across all experimental conditions, the temperature parameter was fixed at 1. This choice preserves the models’ default probability scale, thereby maintaining the interpretability of token-level log-probabilities. 

\subsection{Design}

The study followed a full factorial design in which three factors with two levels each were varied: judgement format, decision type, use of examples (see Figure 1). The levels for each of the factors are explained below.

\begin{figure}[H]
\centering
\includegraphics[width=\linewidth]{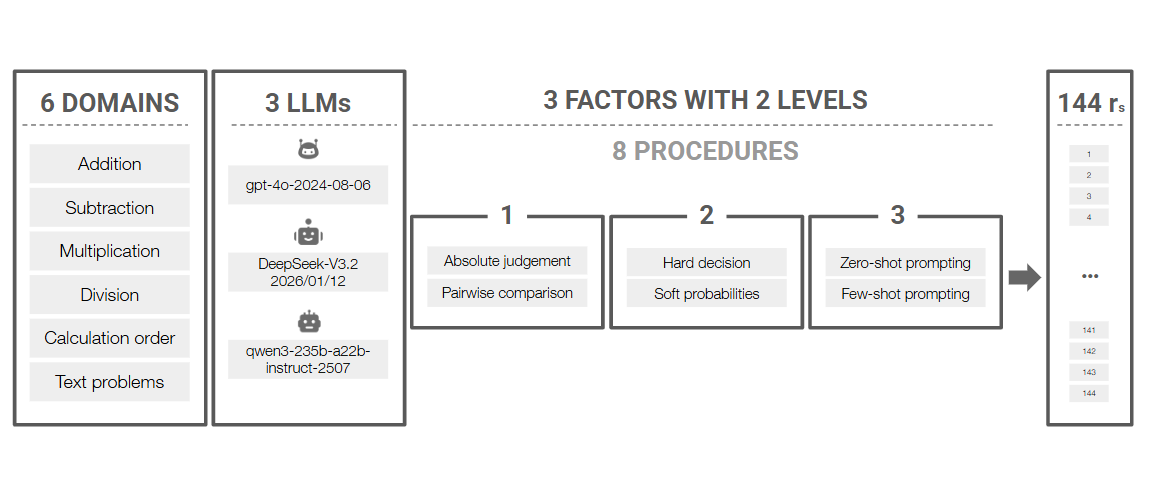}
\caption{Research design scheme}
\label{fig:research_design}
\end{figure}

The central component of this study was the comparison of two judgement formats: absolute judgement and pairwise comparison. Absolute judgement means that each item was presented to an LLM individually, with a request to estimate its absolute difficulty as the expected proportion of target students who would solve it correctly. The responses were provided as a number between 0 and 1 with one decimal place, enclosed in double square brackets, and without any additional commentary (e.g. [[0.5]]).

Pairwise comparison means that items were presented to LLMs in pairs, and the model was asked to predict which item would be more difficult for the target students. The response was a single number: [[1]] if the first item was considred more difficult and [[0]] otherwise. The present study used exhaustive pairing (1,770 pairs per domain) to minimise noise and approximate best-case performance; however, budget-aware designs can reduce the number of required comparisons by sampling a smaller subset of pairs. The order of items within each pair was randomised to mitigate potential order effects.

Each judgement was elicited in a separate dialogue without additional context. The prompts were kept as simple as possible and included only a brief specification of the role to adopt, the intended target audience, task domain, and time allowed for its solution on the platform. No correct answers were provided. The wordings of the tasks were passed to the LLM as raw strings extracted from the dataset. These strings included plain arithmetic expressions, natural-language word problems, HTML formatting. The complete prompt and sample task wordings are reported in Appendix C.

In the hard-decision approach, only the explicit outputs returned by the LLM were used. In the absolute judgement procedure, these outputs were treated as final item difficulty estimates. In the pairwise procedure, they were converted into item difficulty rankings using the Bradley–Terry model (BT; \cite{BradleyTerry1952}).

The soft-probability approach used more than the LLM’s final answer by incorporating the model’s uncertainty. Specifically, it took into account the token-level probabilities assigned to alternative outputs. The number of available token candidates varies across models: 10 for GPT-4o, 20 for DeepSeek-Chat, and 5 for Qwen (Alibaba Cloud)\footnote{When DeepSeek and Qwen are deployed locally, it is possible to access the full set of tokens considered for the response.}. In the absolute judgement procedure, token probabilities were aggregated across candidate numeric outputs spanning 0.0–1.0. In the pairwise procedure, the soft estimate was defined as the probability that the model would return [[1]]. The derivation of soft probability values is provided in Appendix D.

In the zero-shot condition, the LLMs received only the task wording and brief mention of the target audience, assigned role, the time allowed for the task, and task domain. In the few-shot condition, the prompt additionally included two example items paired with their empirical difficulty estimates expressed as a proportion of learners who solved the item correctly. These items were unique for each domain. 

\subsection{Data aggregation and outcome measures}

The LLMs outcomes of the absolute judgement procedure were treated as final difficulty estimates, while the outcomes of the pairwise procedure were transformed with R package BradleyTerry2 \parencite{TurnerFirth2012} for BT model to obtain proxies for item difficulty estimates comparable to IRT estimates.

The agreement with empirical item difficulties was evaluated using Spearman rank correlation to make pairwise and absolute judgement procedures comparable. For each domain, a complete set of correlations across the factorial design (3 × 2 × 2 × 2 = 24 correlations per domain, 144 in total) was calculated. The study examined not only individual correlation levels, but how they vary across factor levels, for example, comparing the average correlation obtained from pairwise-comparison procedures versus absolute judgement procedures.

To understand how different the results would have been if another set of items had been sampled from the original item bank, the bootstrapping procedure was employed. It was used to build confidence intervals for each correlation and for correlation differences between the procedures. Within each domain, items were resampled 10,000 times from 60 items chosen with replacement. For each iteration, the condition-specific results were aggregated across the six domains, and the contrasts between the predefined conditions of interest were computed. The resulting bootstrap distributions were then formed for these contrasts or for condition-specific correlation estimates. Finally, 95\% intervals were obtained from the 2.5th and 97.5th percentiles of the bootstrap values sampled over 10,000 iterations.

\section{Results}
The Results section is structured around the research questions.

\subsection{Overall performance}

\textit{RQ1 Can off-the-shelf LLMs approach the upper bound of difficulty-estimation quality reported in the literature for human raters?} 

Performance aggregated across all experimental conditions (Table 2) showed moderate agreement between LLM-based and empirical difficulty estimates for all three LLMs. Across models, bootstrap confidence intervals overlapped with the range of agreement reported in previous studies for individual human raters \parencite{Clauser2008,MelicanMillsPlake1989,Cross1984}. At the same time, observed correlations were lower than those reported for pooled human judgements, such as the mean correlation of .70 in \textcite{Attali2014}, and lower than the value of .76 reported by \textcite{YanceyEtAl2024} for automated estimation of newly generated word stems.

\begin{table}[t]
\caption{Spearman Correlation Between LLM-Based and Empirical Estimates}
\label{tab:llm_comparison}
\centering
\begin{threeparttable}
\begin{tabular}{@{}l@{\hspace{3 cm}}p{4.0cm}@{}}
\toprule
LLM & Mean [95\% CI] \\
\midrule
GPT-4o        & .673 [.629, .712] \\
DeepSeek Chat & .633 [.584, .675] \\
Qwen 3 235B   & .665 [.616, .709] \\
\bottomrule
\end{tabular}
\begin{tablenotes}[flushleft]
\footnotesize
\item \textit{Note.} Estimates are averaged across domains and design factors. Values in the brackets indicate 95\% bootstrap confidence intervals.
\end{tablenotes}
\end{threeparttable}
\end{table}

\begin{table}[t]
\centering
\caption{Spearman Correlation Between LLM-based and Empirical Difficulties. Averaged across LLMs.}
\label{tab:config_comparison}
\begin{threeparttable}

\begin{tabular}{lllcccccc@{\hspace{20pt}}c}
\toprule
\multicolumn{1}{c}{Configuration} & Add & Sub & Mult & Div & Calc & Text & Mean [95\% CI]  \\


\midrule
Absolute · Hard · Zero-shot & .426 & .582 & .482 & .625 & .551 & .252 & .486 [.430, .540] \\
Absolute · Hard · Few-shot & .632 & .701 & .554 & .646 & .572 & .374 & .580 [.526, .630] \\
Absolute · Soft · Zero-shot & .651 & .753 & .612 & .751 & .602 & .382 & .625 [.571, .673] \\
Absolute · Soft · Few-shot & .740 & .840 & .580 & .738 & .632 & .503 & .672 [.624, .712] \\
\addlinespace
Pairwise · Hard · Zero-shot & .835 & .845 & .678 & .764 & .641 & .617 & .730 [.686, .767] \\
Pairwise · Hard · Few-shot & .814 & .796 & .623 & .762 & .636 & .615 & .708 [.665, .745] \\
Pairwise · Soft · Zero-shot & .835 & .847 & .671 & .771 & .646 & .619 & .732 [.687, .768] \\
Pairwise · Soft · Few-shot & .821 & .807 & .623 & .772 & .665 & .650 & .723 [.681, .759] \\
\bottomrule
\end{tabular}

\begin{tablenotes}[flushleft]
\footnotesize
\item \textit{Note}. Values in the brackets indicate 95\% bootstrap confidence intervals. Add = Addition; Sub = Subtraction; Mult = Multiplication; Div = Division; Calc = Calculation order; Text = Text problems.
\end{tablenotes}
\end{threeparttable}
\end{table}

Table 3 shows the variation in performance across configurations. Higher correlations were observed for pairwise comparisons applied to arithmetic tasks, with some estimates approaching the upper end of the ranges previously reported for human raters. For textual tasks, the corresponding correlations were lower and remained in the moderate range. Absolute judgement combined with token-level probabilities and two examples with known empirical difficulties also produced comparatively strong results.

\subsection{Pairwise comparison versus absolute judgement}

\textit{RQ2 Does the pairwise comparison procedure, as opposed to the absolute judgement, improve the alignment between LLM-derived and empirical item difficulty estimates?} 

\begin{table}[t]
\centering
\caption {Difference in Spearman Correlation Between Pairwise and Absolute Judgement Procedure}
\begin{threeparttable}

\begin{tabular}{lccc}
\toprule
 Configuration & GPT-4o & DeepSeek Chat & Qwen 3 235B \\
\midrule
Average effect & .147* & .122* & .127* \\
Zero-shot · Hard & .275* & .278* & .178* \\
Zero-shot · Soft & .112* & .101* & .105* \\
Few-shot · Hard & .169* & .087* & .128* \\
Few-shot · Soft & .033 & .024 & .096* \\
\bottomrule
\end{tabular}

\begin{tablenotes}[flushleft]
\footnotesize
\item \textit{Note}. Estimates are averaged across domains. An asterisk indicates that the estimate’s 95\% bootstrap confidence interval excludes zero.
\end{tablenotes}
\end{threeparttable}
\end{table}

The bootstrap estimates in Table 4 indicate that the pairwise comparison procedure mostly yields higher Spearman correlations than the absolute judgement procedure. This holds across most prompting configurations, with the exception of few-shot, soft configuration for GPT-4o and DeepSeek Chat - these 95\% bootstrap intervals include zero. The largest gain is observed under the zero-shot, hard configuration. The performance of all LLMs seems relatively similar.

\subsection{Soft vs hard approaches}
\textit{RQ3 Does token-level probability information ('soft' outputs) improve the quality of LLM difficulty estimates compared to 'hard', single-decision outputs?}

\begin{table}[b]
\centering
\caption {Difference in Spearman Correlation Between Soft and Hard Outputs}
\begin{threeparttable}

\begin{tabular}{lccc}
\toprule
Configuration & GPT-4o & DeepSeek Chat & Qwen 3 235B \\
\midrule
Average effect & .086* & .065* & .035* \\
Absolute · Zero-shot & .172* & .171* & .074* \\
Absolute · Few-shot & .150* & .080* & .047* \\
Pairwise · Zero-shot & .008 & -.006 & .002 \\
Pairwise · Few-shot & .014 & .017* & .015* \\
\bottomrule
\end{tabular}

\begin{tablenotes}[flushleft]
\footnotesize
\item \textit{Note}.  Estimates are averaged across domains. An asterisk indicates that the estimate’s 95\% bootstrap confidence interval excludes zero.
\end{tablenotes}
\end{threeparttable}
\end{table}

Table~5 reports bootstrap estimates of the difference between the soft-probability approach and the hard-decision approach ($\Delta=r_{s,\text{soft}}-r_{s,\text{hard}}$), aggregated across domains. The soft approach yields a consistent improvement for the absolute procedure for all three LLMs with the advantage of the zero-shot strategy. In contrast, within the pairwise procedure, the soft--hard differences are small and often include zero, indicating little systematic benefit of soft probabilities for pairwise comparisons. 

\subsection{Zero-shot vs few-shot strategies}

\textit{RQ4 How does a few-shot prompting affect the quality of LLM difficulty estimates compared to a zero-shot strategy?}

\begin{table}[t]
\centering
\caption {Difference in Spearman Correlation Between Few-shot and Zero-shot Outputs} 
\begin{threeparttable}

\begin{tabular}{lccc}
\toprule
Configuration & GPT-4o & DeepSeek Chat & Qwen 3 235B \\
\midrule
Average effect & .046* & .055* & -.019 \\
Absolute · Hard & .103* & .167* & .010 \\
Absolute · Soft & .081* & .077* & -.017 \\
Pairwise · Hard & -.003 & -.024 & -.040* \\
Pairwise · Soft & .002 & -.001 & -.027 \\
\bottomrule
\end{tabular}

\begin{tablenotes}[flushleft]
\footnotesize
\item \textit{Note}. Estimates are averaged across domains. An asterisk indicates that the estimate’s 95\% bootstrap confidence interval excludes zero.
\end{tablenotes}
\end{threeparttable}
\end{table}

Table~6 summarises bootstrap estimates of the prompting effect defined as ($\Delta=r_{s,\text{few-shot}}-r_{s,\text{zero-shot}}$), aggregated across domains. On average, few-shot prompting yields a small but stable improvement for GPT-4o and DeepSeek Chat, while Qwen~3 does not show clear benefit. The positive few-shot advantage for GPT-4o and DeepSeek is driven by the absolute judgement procedure. In contrast, pairwise comparisons show little to no benefit from few-shot prompting: mean estimates are close to or below zero and intervals often include zero. 

\begin{table}[!htbp]
\centering
\caption {Spearman correlation aggregated across conditions and LLMs}
\begin{threeparttable}

\begin{tabular}{l r}
\toprule
Domain & Mean [95\% CI] \\
\midrule
Addition & .719 [.635, .785] \\
Subtraction & .771 [.711, .817] \\
Multiplication & .603 [.503, .698] \\
Division & .729 [.645, .795] \\
Calculation order & .618 [.483, .727] \\
Text problems & .501 [.357, .628] \\
\bottomrule
\end{tabular}

\begin{tablenotes}[flushleft]
\footnotesize
\item \textit{Note}. Values in the brackets indicate the bounds of the 95\% bootstrap confidence intervals.
\end{tablenotes}
\end{threeparttable}
\end{table}

\subsection{Variability across domains}

\textit{RQ5 How much does the quality of LLM difficulty estimates vary across different domains?}

Across domains, performance is consistently positive, but clearly heterogeneous. The two basic arithmetic domains — Addition and Subtraction — show the strongest agreement with empirical difficulty, with point estimates clustered around the upper range (roughly 0.7–0.8) and comparatively tight intervals, suggesting stable performance across bootstrap resamples. Division is similarly high, though slightly below the very top performers. A noticeable decline is observed for Multiplication and Calculation Order, which fall in the mid-range (approximately 0.6–0.7) and show a greater uncertainty, indicating greater variability in difficulty recovery. The weakest results are found for Text Problems, where the mean correlation is lowest (around 0.5). Because the Text Problems items are presented in Dutch, it remains unclear whether this pattern reflects greater structural complexity, language effect, or both.

\section{Conclusions and discussion}

This study examined whether LLMs can serve as scalable judges for item difficulty estimation and how different elicitation procedures affect their alignment with empirical difficulty. Across models, LLM-derived estimates demonstrated moderate-to-strong rank-order agreement with empirical difficulty values, suggesting that LLMs can provide meaningful information for preliminary item calibration when pretest data are limited or absent. 

Two methodological patterns emerged. First, pairwise comparisons generally produced stronger alignment than absolute judgements, with the largest advantages observed under zero-shot, hard-decision settings. This result is in line with earlier claims in the comparative judgement literature, where relative judgements have often been said to be of higher quality than direct absolute estimation. However, because pairwise strategies are more intense in terms of time and financial costs (see Appendix E), sometimes absolute judgement procedures could be considered if they are combined with the soft probability approach and a few examples are available. Second, the use of soft probability information tended to improve performance relative to hard decisions primarily within the absolute judgement procedure. Domain analysis indicated that performance depends on the task domain, with more complex domains, such as Text Problems, remaining more difficult to estimate reliably than basic arithmetic ones. The use of examples in prompting showed little or no benefit, with effects concentrated mainly in the absolute judgement, hard-decision condition. This contrasts with the popularity of few-shot prompting in other areas and suggests a possible direction for future research.

In addition, a few other limitations of the present study point to ideas that could be interesting to future researchers. 

First, the present study does not examine the absolute differences between LLM-based estimates and those derived from student response data, which limits the immediate applicability of the findings. This is because estimates produced by pairwise comparison are located on a relative scale. A useful next step would be to incorporate anchor items with known difficulty values into the Bradley–Terry procedure and then assess the agreement between empirical and LLM-based estimates using Pearson correlation and error-based metrics such as mean squared error (MSE).

Second, all judgements were collected without persistent context, so each LLM decision was made independently of the previous ones. This design allowed for a clean comparison of procedures, but it may also have constrained performance, particularly for items that are close in difficulty and for absolute judgement settings, where calibration is especially challenging. Future work could investigate more context-rich designs, including the comparison of triplets or larger item sets, the presentation of items in batches, and the use of a longer conversational context in which previous judgements are retained and can inform subsequent decisions.

Third, performance can potentially be improved through more explicit cross-model collaboration. This could range from relatively simple aggregation procedures to controlled ensembles or more structured multi-agent discussion formats. Likewise, other LLM configurations may prove beneficial, including multi-step reasoning procedures, the provision of correct answers, and a more systematic search for prompt wordings.

A further limitation concerns interpretability. The present design does not explain why particular items were judged to be difficult or easy. Future research could examine which item features are associated with difficulty, such as the concepts required, the number of reasoning steps, or linguistic complexity. This work could combine NLP-based feature extraction with expert ratings and interviews that make human judgement processes more explicit.

Overall, the present results support the view that LLMs can contribute to item difficulty estimation in cold-start or low-data settings, but also suggest that this use is best understood as an emerging methodological direction rather than as a finished solution. Further progress will likely depend not only on stronger models, but also on better judgement designs, richer use of context, and more interpretable accounts of what drives perceived item difficulty.


\section{Generative AI use statement}
Large language models were used in this study as the object of investigation: their outputs formed part of the empirical material analysed in the paper. In addition to that, GPT-4o was used in a limited manner during manuscript preparation for language editing. All methodological decisions, analyses, interpretations and final text were suggested and approved by the authors.

\section{Appendices}

\subsection{Appendix A. Examples of the tasks}

The screenshots contained in Figure 2 show how the children see the tasks from different games (Subtraction, Division and Text Problems) on the website of Prowise Learn.

\begin{figure}[htbp]
    \centering
    \includegraphics[width=0.3\textwidth]{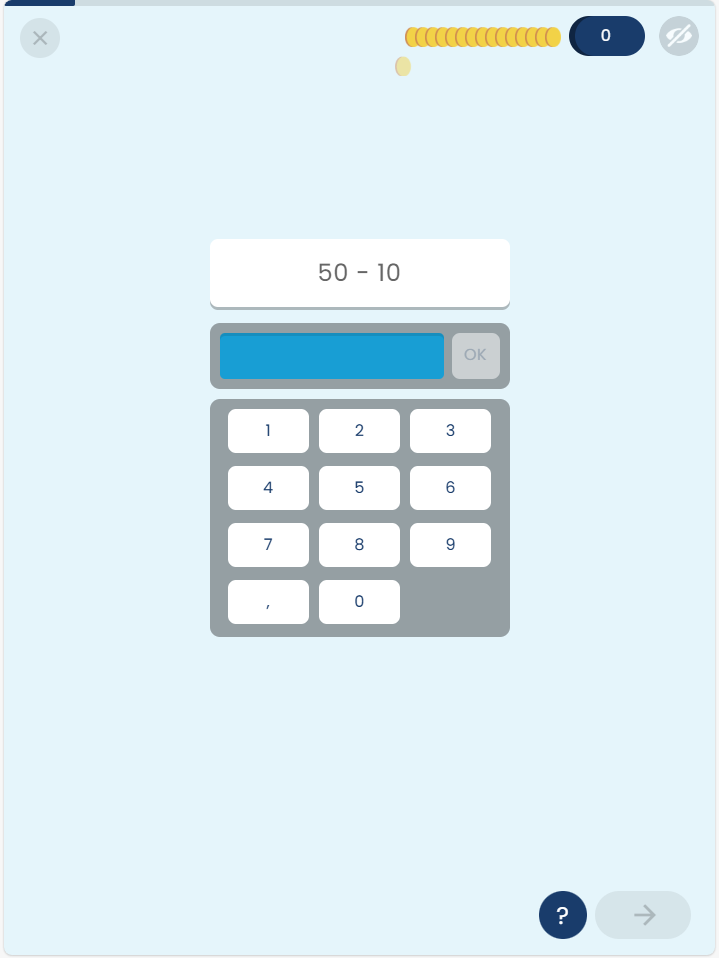}
    \hfill
    \includegraphics[width=0.3\textwidth]{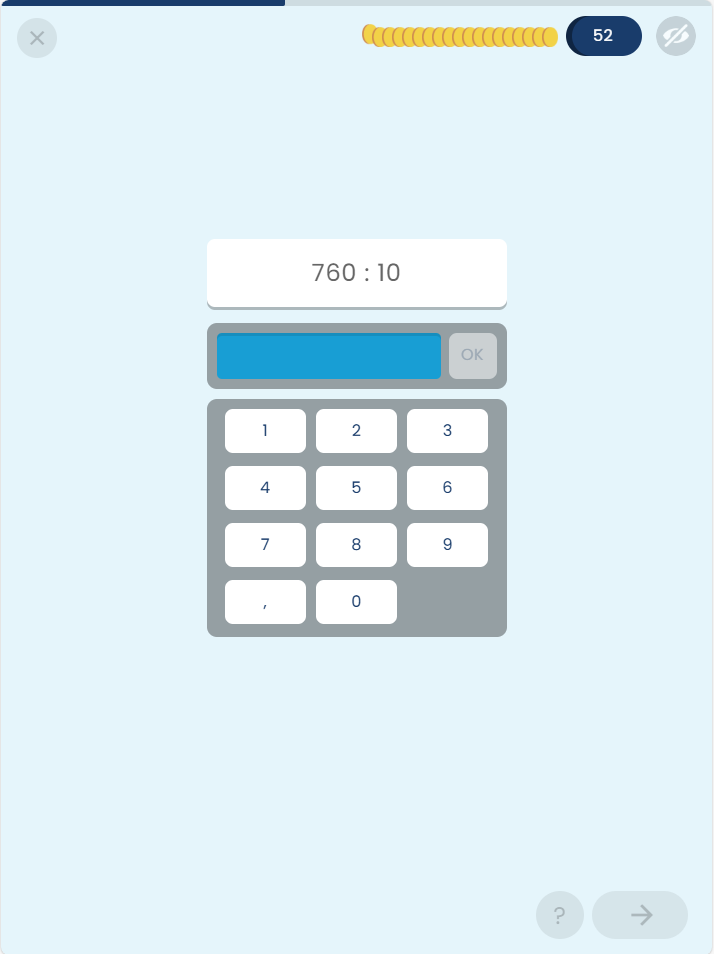}
    \hfill
    \includegraphics[width=0.3\textwidth]{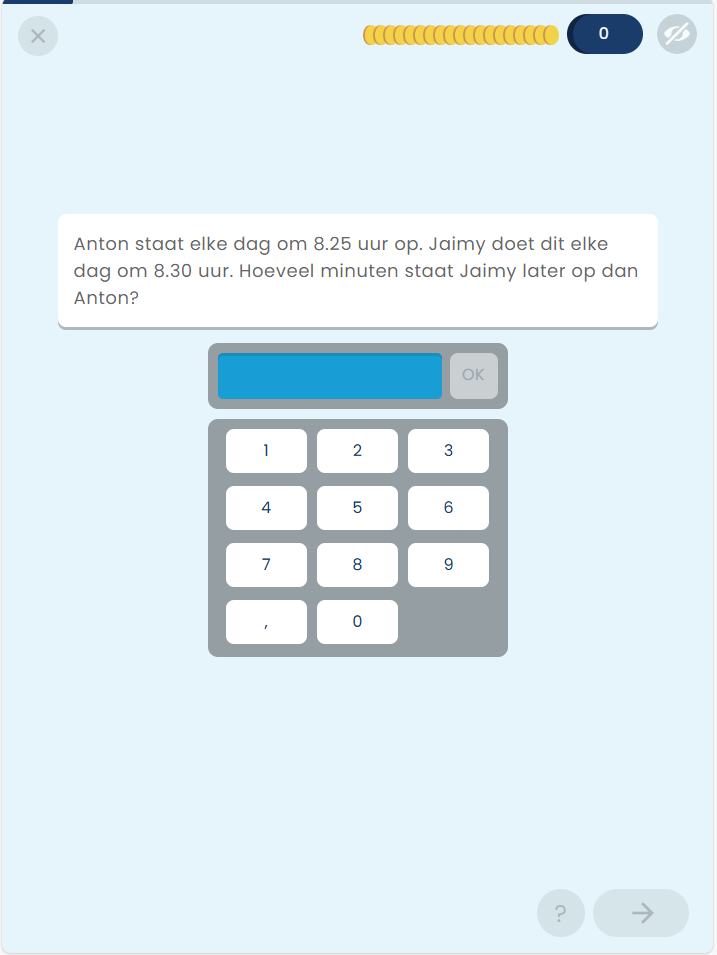}
    \caption{Prowise Learn Tasks Examples}
\end{figure}

\subsection{Appendix B. Difficulty estimation in the dataset}

It should be noted that the original data were obtained from an adaptive learning system, which means that items were assigned to students based on their estimated ability level changing over time. To estimate difficulty for this setting, the response log data were first preprocessed to define the response vectors with the student’s responses within a single session where approximate stability of ability could be assumed. If an item was repeated within a session, only the first attempt was retained, and sessions with fewer than five unique items were excluded. Because users could contribute multiple sessions, person weights were applied, calculated as $w = \frac{1}{S}$, where $S$ is the number of sessions, so that each user contributed approximately equally to the estimation. 

The sessions were then recoded and treated as separate users. Item difficulties and the standard deviation of ability in the population parameters were estimated with a custom routine that implements marginal maximum likelihood via the Expectation–Maximisation (EM) algorithm. The E-step uses Gauss--Hermite quadrature (60 nodes) to approximate integrals over the latent ability distribution, and the M-step updates item parameters iteratively until convergence (tolerance $10^{-5}$; up to 100 EM iterations).

The algorithm returned item parameter estimates and an estimate of the standard deviation of student ability. The estimated item parameters corresponded to (minus) item difficulty on the logit scale. Finally, for each item CTT expected proportion correct was computed by integrating the Rasch logistic response function over the estimated normal ability distribution:
\[
p_i \;=\; \int_{-\infty}^{\infty} \frac{\exp(\theta-b_i)}{1+\exp(\theta-b_i)} \, \phi(\theta;0,\sigma)\, d\theta,
\]
where $b_i$ is the Rasch item difficulty and $\sigma$ is the estimated standard deviation of ability for that domain. This integral was evaluated numerically with bounds $[-50,50]$, yielding an expected $p$-value for each item, which was saved together with the logit-scale difficulty estimate.

\subsection{Appendix C. Full texts of prompts}

\paragraph{Prompt A (absolute judgement; zero-shot).}
\begin{quote}\small\ttfamily
\textbf{System message.}\\
You are an expert in primary education and educational assessment.\\
You help to evaluate the difficulty of the tasks for Dutch students in grade <GRADE>.\\
All tasks are about mathematics, in particular <DOMAIN>. You will receive one task and must estimate its difficulty.\\
Real students have at most <TIME\_LIMIT> seconds to answer it.\\
Base your judgment on the syntax, logic, and required understanding.\\[0.75em]
\textbf{User message.}\\
Guess the proportion of students who solved this task correctly.\\
Provide the answer as a number between 0 and 1 with one decimal place.\\
Write it in brackets like this [[0.5]], don't provide any verbal explanations\\[0.75em]
<TASK\_TEXT>
\end{quote}

\paragraph{Prompt B (absolute judgement; few-shot).}
\begin{quote}\small\ttfamily
\textbf{System message.}\\
You are an expert in primary education and educational assessment.\\
You help to evaluate the difficulty of the tasks for Dutch students in grade <GRADE>.\\
All tasks are about mathematics, in particular <DOMAIN>. You will receive one task and must estimate its difficulty.\\
Real students have at most <TIME\_LIMIT> seconds to answer it.\\
Base your judgment on the syntax, logic, and required understanding.\\[0.75em]
Here are two example tasks with known difficulty rates:\\[0.5em]
--- Example Task 1 ---\\
<EXAMPLE\_TASK\_1\_TEXT>\\
**Difficulty:** <EXAMPLE\_1\_DIFFICULTY\_\%> of students solved this correctly on the first try.\\[0.75em]
--- Example Task 2 ---\\
<EXAMPLE\_TASK\_2\_TEXT>\\
**Difficulty:** <EXAMPLE\_2\_DIFFICULTY\_\%> of students solved this correctly on the first try.\\[0.75em]
\textbf{User message.}\\
Guess the proportion of students who solved this task correctly.\\
Provide the answer as a number between 0 and 1 with one decimal place.\\
Write it in brackets like this [[0.5]], don't provide any verbal explanations. Here is the task.\\[0.75em]
<TASK\_TEXT>
\end{quote}

\paragraph{Prompt C (pairwise comparison; zero-shot).}
\begin{quote}\small\ttfamily
\textbf{System message.}\\
You are an expert in primary education and educational assessment.\\
You help to evaluate the difficulty of the tasks for Dutch students in the grade <GRADE>.\\
All tasks are about mathematics, in particular <DOMAIN>.\\
You will receive two tasks and must decide which one is more difficult for the average student.\\
Real students have at most <TIME\_LIMIT> seconds to answer it.\\
Base your judgment on the syntax, logic, and required understanding.\\[0.75em]
\textbf{User message.}\\
Now compare the two tasks below.\\[0.5em]
--- Task A ---\\
<TASK\_A\_TEXT>\\[0.75em]
--- Task B ---\\
<TASK\_B\_TEXT>\\[0.75em]
Which task is more difficult for an average student?\\
Write [[1]] if Task A is more difficult.\\
Write [[0]] if Task B is more difficult.\\
Respond only with [[1]] or [[0]]. Make your best guess.
\end{quote}

\paragraph{Prompt D (pairwise comparison; few-shot).}
\begin{quote}\small\ttfamily
\textbf{System message.}\\
You are an expert in primary education and educational assessment.\\
You help to evaluate the difficulty of the tasks for Dutch students in the grade <GRADE>.\\
All tasks are about mathematics, in particular <DOMAIN>.\\
You will receive two tasks and must decide which one is more difficult for the average student.\\
Real students have at most <TIME\_LIMIT> seconds to answer it.\\
Base your judgment on the syntax, logic, and required understanding.\\[0.75em]
Here are two example tasks with known difficulty rates:\\[0.5em]
--- Example Task 1 ---\\
<EXAMPLE\_TASK\_1\_TEXT>\\
**Difficulty:** <EXAMPLE\_1\_DIFFICULTY\_\%> of students solved this correctly on the first try.\\[0.75em]
--- Example Task 2 ---\\
<EXAMPLE\_TASK\_2\_TEXT>\\
**Difficulty:** <EXAMPLE\_2\_DIFFICULTY\_\%> of students solved this correctly on the first try.\\[0.75em]
\textbf{User message.}\\
Now compare the two tasks below.\\[0.5em]
--- Task A ---\\
<TASK\_A\_TEXT>\\[0.75em]
--- Task B ---\\
<TASK\_B\_TEXT>\\[0.75em]
Which task is more difficult for an average student?\\
Write [[1]] if Task A is more difficult.\\
Write [[0]] if Task B is more difficult.\\
Respond only with [[1]] or [[0]]. Make your best guess.
\end{quote}

\paragraph{Task wordings examples used in the prompts.}

\begin{enumerate}
\item Addition: \verb |"0 + 18"|.
\item Subtraction: \verb |"5400 - 4000"|.
\item Multiplication: \verb |"20 x 4"|.                   
\item Division: \verb |"30 : 0,6<br><small>denk aan 30 : 6 = 5</small>"|.   
\item Calculation Order: 
\verb|"$3\\times5+3\\times4$"|.
\item Text Problems: \texttt{\detokenize{"In de nieuwste Donald Duck zitten 28 pagina's en in de nieuwste Tina ook. Hoeveel pagina's zijn er in totaal als Armin en zijn zusje allebei een tijdschrift krijgen?"}}
\end{enumerate}

\subsection{Appendix D. Working with soft probabilities}

\subsubsection{Absolute judgement case}
In the absolute judgement procedure, the model outputs a numeric judgement on a bounded
scale from 0.0 to 1.0. Because the numeric output is represented by three tokens, one needs to distinguish two cases.

In case A the output is contained in the interval $[0,1)$. It is typically tokenised into three tokens (``0'', ``.'', and a digit). At each required position, the set of candidate tokens returned by the API (up to a deployment-specific top-$k$ limit) is taken from the token-level log-probabilities and renormalised over the subset of relevant tokens---\{0,1\} for the leading integer token and $\{0,\ldots,9\}$ for the fractional digit---so that the probabilities sum to 1 within each subset. The soft estimate used in this work combines the uncertainty about the leading integer token $I\in\{0,1\}$ with the uncertainty about the fractional digit $D\in\{0,\ldots,9\}$. It is computed as
\[
\hat p_{\text{soft}} = P(I=1) + P(I=0)\,\mathrm{EV}_{\text{frac}},
\qquad
\mathrm{EV}_{\text{frac}}=\sum_{d=0}^{9}(d/10)\,P(D=d),
\]
where $P(I=\cdot)$ and $P(D=\cdot)$ denote the renormalised token probabilities at the corresponding generation positions.

In case B the parsed output is equal to 1 (i.e., \texttt{[[1.0]]}). This outcome provides limited information on the competing values in $[0,1)$, because the subsequent token-level probabilities observed under the \texttt{1.0} branch do not characterise the fractional-digit uncertainty relevant for \texttt{0.$d$}. To handle this case, the procedure records the renormalised distribution over the leading integer token $I\in\{0,1\}$ from the first API call (yielding $P(I=1)$ and $P(I=0)$).

If the first call returns \texttt{[[1.0]]} and the model is highly confident in the leading token being \texttt{``1''} (operationalised as $P(I=1)\ge 0.99$), the algorithm does not resample; instead, it computes a conservative soft estimate using a fixed fractional expectation of $0.9$:
\[
\hat p_{\text{soft}} = P(I=1) + P(I=0)\cdot 0.9.
\]

Otherwise, the same request is repeated (up to a maximum of 100 attempts) until a completion of the form \texttt{[[0.$d$]]} is obtained. Once a \texttt{0.$d$} completion is observed, the fractional-digit distribution $P(D=d)$ is estimated from the renormalised token probabilities at the fractional position, yielding
\[
\mathrm{EV}_{\text{frac}}=\sum_{d=0}^{9}(d/10)\,P(D=d),
\]
and the final soft estimate is computed as in Case~A:
\[
\hat p_{\text{soft}} = P(I=1) + P(I=0)\,\mathrm{EV}_{\text{frac}}.
\]
If resampling reaches the maximum number of attempts without observing a \texttt{0.$d$} completion, the algorithm falls back to the first-call integer-token probabilities and again sets $\mathrm{EV}_{\text{frac}}=0.9$, i.e., $\hat p_{\text{soft}} = P(I=1) + P(I=0)\cdot 0.9$.

\subsubsection{Pairwise comparison task}
\paragraph{Deriving soft probabilities for the pairwise procedure.}
For the pairwise elicitation, each comparison presents two items $(i,j)$ and the model returns a binary preference indicating which item is harder. In addition to the sampled response, the API returns a truncated top-$k$ list of candidate tokens with token-level log-probabilities at the decision position. For the \emph{soft} variant, these log-probabilities are converted into a probability that item $i$ (option 1) is harder versus item $j$ (option 2). The probabilities are normalised over the relevant token set $\{\texttt{``1''},\texttt{``0''}\}$.
These soft probabilities are interpreted as fractional ``win'' masses for the Bradley--Terry likelihood, where token \texttt{``1''} corresponds to item $i$ being harder, and token \texttt{``0''} corresponds to item $j$ being harder. To use the standard Bradley--Terry implementation with count data, the fractional masses are converted to pseudo-counts by rounding to 6 digits and scaling with a constant of $10^6$. The resulting matrix of pseudo-counts is then passed to the Bradley--Terry model to estimate item abilities on the latent difficulty scale.

\begin{table}[t]
\centering
\caption {Time spent on API requests to GPT-4o. Configurations comparison}
\begin{tabular}{lcccc}
\toprule
 & \begin{tabular}[c]{@{}c@{}}Absolute\\ Zero-shot\end{tabular}
 & \begin{tabular}[c]{@{}c@{}}Absolute\\ Few-shot\end{tabular}
 & \begin{tabular}[c]{@{}c@{}}Pairwise\\ Zero-shot\end{tabular}
 & \begin{tabular}[c]{@{}c@{}}Pairwise\\ Few-shot\end{tabular} \\
\midrule
Number of judgements        & 360 & 360 & 7020 & 7020 \\
Total time (sec)            & 316 & 248 & 6943 & 6961 \\

\midrule
\end{tabular}
\end{table}

\begin{table}[t]
\centering
\caption {Tokens and money spent on API requests to GPT-4o. Configurations comparison}
\begin{tabular}{lcccccccc}
\toprule
 & \multicolumn{2}{c}{Absolute Zero-shot} & \multicolumn{2}{c}{Absolute Few-shot} & \multicolumn{2}{c}{Pairwise Zero-shot} & \multicolumn{2}{c}{Pairwise Few-shot} \\
\cmidrule(lr){2-3}\cmidrule(lr){4-5}\cmidrule(lr){6-7}\cmidrule(lr){8-9}
All domains & Tokens & \$ & Tokens & \$ & Tokens & \$ & Tokens & \$ \\
\midrule
Prompt tokens     & 57,400   & 0.14 & 83,355   & 0.21 & 2,002,401 & 5.01 & 2,828,991 & 7.07 \\
Completion tokens & 1,990    & 0.02 & 1,800    & 0.02 & 31,860    & 0.32 & 31,860    & 0.32 \\
Total             & 59,390   & 0.16 & 85,155   & 0.23 & 2,034,261 & 5.32 & 2,860,851 & 7.39 \\
\bottomrule
\end{tabular}
\end{table}

\subsection{Appendix E. Token, money and time costs.}

Tables 8 and 9 summarise the computational costs of the different procedures for GPT-4o, with Table 8 reporting API request time and Table 9 reporting token usage and corresponding financial cost. Other LLMs took roughly two to three times longer to complete comparable procedures with approximately the same token usage. Although the pairwise procedure could be optimised to some extent \parencite{Crompvoets2020, RainaLiusieGales2025}, it would still require a substantially larger number of requests, which implies being less economical in token and time usage.

\printbibliography

\end{document}